\documentclass[preprintnumbers,amsmath,amssymbm,prd]{revtex4}
\usepackage{epsfig}
\usepackage{graphicx}
\usepackage{amssymb}

\begin{document}
\title{Quasi-bound states of massive scalar fields in the Kerr black-hole spacetime: Beyond the hydrogenic approximation}
\author{Shahar Hod}
\affiliation{The Ruppin Academic Center, Emeq Hefer 40250, Israel}
\affiliation{ } \affiliation{The Hadassah Institute, Jerusalem
91010, Israel}
\date{\today}

\begin{abstract}
\ \ \ Rotating black holes can support quasi-stationary (unstable)
bound-state resonances of massive scalar fields in their exterior
regions. These spatially regular scalar configurations are
characterized by instability timescales which are much longer than
the timescale $M$ set by the geometric size (mass) of the central
black hole. It is well-known that, in the small-mass limit
$\alpha\equiv M\mu\ll1$ (here $\mu$ is the mass of the scalar
field), these quasi-stationary scalar resonances are characterized
by the familiar {\it hydrogenic} oscillation spectrum:
$\omega_{\text{R}}/\mu=1-\alpha^2/2{\bar n}^2_0$, where the integer
$\bar n_0(l,n;\alpha\to0)=l+n+1$ is the principal quantum number of
the bound-state resonance (here the integers $l=1,2,3,...$ and
$n=0,1,2,...$ are the spheroidal harmonic index and the resonance
parameter of the field mode, respectively). As it depends only on
the principal resonance parameter $\bar n_0$, this {\it small}-mass
($\alpha\ll1$) hydrogenic spectrum is obviously {\it degenerate}. In
this paper we go beyond the small-mass approximation and analyze the
quasi-stationary bound-state resonances of massive scalar fields in
rapidly-spinning Kerr black-hole spacetimes in the regime
$\alpha=O(1)$. In particular, we derive the non-hydrogenic (and, in
general, {\it non}-degenerate) resonance oscillation spectrum
${{\omega_{\text{R}}}/{\mu}}=\sqrt{1-(\alpha/{\bar n})^2}$,
where $\bar n(l,n;\alpha)=\sqrt{(l+1/2)^2-2m\alpha+2\alpha^2}+1/2+n$
is the {\it generalized} principal quantum number of the
quasi-stationary resonances.
This analytically derived formula for the characteristic oscillation
frequencies of the composed black-hole-massive-scalar-field system
is shown to agree with direct numerical computations of the
quasi-stationary bound-state resonances.
\end{abstract}
\bigskip
\maketitle


\section{Introduction}

Recent analytical \cite{Hodrc} and numerical \cite{HerR} studies of
the coupled Einstein-scalar equations have revealed that rotating
black holes can support stationary spatially regular configurations
of {\it massive} scalar fields in their exterior regions. These
bound-state resonances of the composed black-hole-scalar-field
system owe their existence to the well-known phenomenon of
superradiant scattering \cite{Zel,PressTeu1} of integer-spin
(bosonic) fields in rotating black-hole spacetimes.

The stationary black-hole-scalar-field configurations
\cite{Hodrc,HerR} mark the boundary between stable and unstable
bound-state resonances of the composed system. In particular, these
stationary scalar field configurations are characterized by
azimuthal frequencies $\omega_{\text{field}}$ which are in resonance
with the black-hole angular velocity $\Omega_{\text{H}}$
\cite{Noteunits}:
\begin{equation}\label{Eq1}
\omega_{\text{field}}=m\Omega_{\text{H}}\  ,
\end{equation}
where $m=1,2,3,...$ is the azimuthal quantum number of the field
mode. Bound-state field configurations in the superradiant regime
$\omega_{\text{field}}<m\Omega_{\text{H}}$ are known to be unstable
(that is, grow in time), whereas bound-state field configurations in
the regime $\omega_{\text{field}}>m\Omega_{\text{H}}$ are known to
be stable (that is, decay in time) \cite{Zel,PressTeu1}.

The bound-state scalar resonances of the rotating Kerr black-hole
spacetime are characterized by at least two different time scales:
(1) the typical oscillation period $\tau_{\text{oscillation}}\equiv
2\pi/\omega_{\text{R}}\sim 1/\mu$ of the bound-state massive scalar
configuration (here $\mu$ is the mass of the scalar field
\cite{Notedim}), and (2) the instability growth time scale
$\tau_{\text{instability}}\equiv 1/\omega_{\text{I}}$ associated
with the superradiance phenomenon. Former studies
\cite{Det,Zour,Dol1,HodSO,Bri} of the Einstein-massive-scalar-field
system have revealed that these two time scales are well separated.
In particular, it was shown \cite{Det,Zour,Dol1,HodSO,Bri} that the
composed system is characterized by the relation
\begin{equation}\label{Eq2}
\tau_{\text{instability}}\gg \tau_{\text{oscillation}}\  ,
\end{equation}
or equivalently
\begin{equation}\label{Eq3}
\omega_{\text{I}}\ll \omega_{\text{R}}\  .
\end{equation}
The strong inequality (\ref{Eq2}) implies that the bound-state
massive scalar configurations may be regarded as the {\it
quasi-stationary} resonances of the composed system.

As shown in \cite{Dol2,CarS}, the physical significance of the
characteristic black-hole-scalar-field oscillation frequencies
$\{\omega_{\text{R}}(n)\}_{n=0}^{n=\infty}$ \cite{Notenn} lies in
the fact that the corresponding quasi-stationary scalar resonances
dominate the dynamics of massive scalar fields in curved black-hole
spacetimes. In particular, recent numerical simulations
\cite{Dol2,CarS} of the dynamics of massive scalar fields in the
Kerr black-hole spacetime have demonstrated explicitly that these
quasi-stationary bound-state resonances dominate the characteristic
Fourier power spectra $P(\omega)$ of the composed
black-hole-massive-scalar-field system \cite{Notefor}.

\section{The small-mass hydrogenic spectrum}

As shown by Detweiler \cite{Det}, the massive scalar resonances can
be calculated analytically in the small-mass regime $M\mu\ll1$. In
particular, one finds \cite{Det} that the quasi-stationary
bound-state scalar resonances are characterized by the familiar {\it
hydrogenic} spectrum
\begin{equation}\label{Eq4}
{{\omega_{\text{R}}(\bar n_0)}\over{\mu}}=1-{{\alpha^2}\over{2{\bar
n}^2_0}}\ \ \ \ \text{for}\ \ \ \alpha\equiv M\mu\ll1\  ,
\end{equation}
where the integer
\begin{equation}\label{Eq5}
\bar n_0(l,n;\alpha\to0)=l+1+n
\end{equation}
is the principal quantum number of the quasi bound-state resonances.
Here the integer $l\geq|m|$ is the spheroidal harmonic index of the
field mode and $n=0,1,2,...$ is the resonance parameter.

It is worth emphasizing that the hydrogenic spectrum (\ref{Eq4})
depends only on the principal resonance parameter (quantum number)
$\bar n_0=l+1+n$. This small-mass oscillation spectrum is therefore
{\it degenerate}. That is, two different modes, $(l,n)$ and
$(l',n')$ with $l+n=l'+n'$, are characterized by the {\it same}
resonant frequency:
$\omega_{\text{R}}(l,n)=\omega_{\text{R}}(l',n')$ for $l+n=l'+n'$.

To the best of our knowledge, the oscillation frequency spectrum
which characterizes the quasi-stationary bound-state resonances of
massive scalar fields in the rotating Kerr black-hole spacetime has
not been studied analytically beyond the hydrogenic regime
(\ref{Eq4}) of small ($\alpha\ll1$) field masses. The main goal of
the present paper is to analyze the oscillation spectrum of the
composed black-hole-massive-scalar-field system in the $\alpha=O(1)$
regime. To that end, we shall use the resonance equation [see Eq.
(\ref{Eq11}) below] derived in \cite{HodSO} for the bound-state
resonances of massive scalar fields in rapidly-rotating
(near-extremal) Kerr black-hole spacetimes. As we shall show below,
this resonance equation can be solved {\it analytically} to yield
the characteristic oscillation spectrum
$\{\omega_{\text{R}}(n)\}_{n=0}^{n=\infty}$ of the quasi-stationary
bound-state scalar resonances in the regime $\alpha\lesssim1$.

\section{Description of the system}

We consider a scalar field $\Psi$ of mass $\mu$ linearly coupled to
a rapidly-rotating (near-extremal) Kerr black hole of mass $M$ and
dimensionless angular momentum $a/M\to 1^-$. The dynamics of the
scalar field in the black-hole spacetime is governed by the
Klein-Gordon (Teukolsky) wave equation
\begin{equation}\label{Eq6}
(\nabla^a \nabla_a -\mu^2)\Psi=0\  .
\end{equation}
Substituting the field decomposition \cite{Notebo,Kerr,Notemn}
\begin{equation}\label{Eq7}
\Psi(t,r,\theta,\phi)=\int\sum_{l,m}
e^{im\phi}S_{lm}(\theta)R_{lm}(r)e^{-i\omega t}d\omega\
\end{equation}
into the wave equation (\ref{Eq6}), one finds \cite{Teuk,Stro} that
the radial function $R$ and the angular function $S$ obey two
ordinary differential equations of the confluent Heun type
\cite{Heun,Fiz1}.

The angular eigenfunctions, known as the spheroidal harmonics, are
determined by the angular Teukolsky equation
\cite{Teuk,Stro,Heun,Fiz1,Abram}
\begin{equation}\label{Eq8}
{1\over {\sin\theta}}{\partial \over
{\partial\theta}}\Big(\sin\theta {{\partial
S_{lm}}\over{\partial\theta}}\Big)+\Big[a^2(\omega^2-\mu^2)\cos^2\theta-{{m^2}\over{\sin^2\theta}}+A_{lm}\Big]S_{lm}=0\
.
\end{equation}
The regularity requirements of these functions at the two boundaries
$\theta=0$ and $\theta=\pi$ single out a discrete set of angular
eigenvalues $\{A_{lm}\}$ [see Eq. (\ref{Eq14}) below] labeled by the
integers $l$ and $m$.

The radial Teukolsky equation is given by \cite{Teuk,Stro,Hodcen}
\begin{equation}\label{Eq9}
\Delta{{d}
\over{dr}}\Big(\Delta{{dR_{lm}}\over{dr}}\Big)+\Big\{[(r^2+a^2)\omega-am]^2
-\Delta(a^2\omega^2-2ma\omega+\mu^2r^2+A_{lm})\Big\}R_{lm}=0\ ,
\end{equation}
where $\Delta\equiv (r-r_+)(r-r_-)$ \cite{Notepm}. Note that the
angular Teukolsky equation (\ref{Eq8}) and the radial Teukolsky
equation (\ref{Eq9}) are coupled by the angular eigenvalues
$\{A_{lm}\}$.

The quasi-stationary bound-state resonances of the massive scalar
fields in the black-hole spacetime are characterized by the boundary
conditions of purely ingoing waves at the black-hole horizon (as
measured by a comoving observer) and a spatially decaying (bounded)
radial eigenfunction at asymptotic infinity
\cite{Det,Dol1,Zour,Notetor}:
\begin{equation}\label{Eq10}
R_{lm} \sim
\begin{cases}
{1\over r}e^{-\sqrt{\mu^2-\omega^2}y} & \text{ as }
r\rightarrow\infty\ \ (y\rightarrow \infty)\ ; \\
e^{-i (\omega-m\Omega_{\text{H}})y} & \text{ as } r\rightarrow r_+\
\ (y\rightarrow -\infty)\ ,
\end{cases}
\end{equation}
where $\Omega_{\text{H}}$ is the angular velocity of the black-hole
horizon [see Eq. (\ref{Eq12}) below]. The boundary conditions
(\ref{Eq10}) imposed on the radial eigenfunctions single out a
discrete set of eigenfrequencies
$\{\omega_n(a/M,l,m,\alpha)\}_{n=0}^{n=\infty}$ which characterize
the quasi-stationary bound-state resonances of the massive scalar
fields in the Kerr black-hole spacetime \cite{Det,Dol1,Zour}.

\section{The characteristic resonance equation and its regime of validity}

Solving analytically the radial Klein-Gordon (Teukolsky) equation
(\ref{Eq9}) in two different asymptotic regions and using a standard
matching procedure for these two radial solutions in their common
overlap region [see Eq. (\ref{Eq15}) below], we have derived in
\cite{HodSO} the characteristic resonance equation
\begin{equation}\label{Eq11}
{1\over{\Gamma({1\over
2}+\beta-\kappa)}}=\Big[{{\Gamma(-2\beta)}\over{\Gamma(2\beta)}}\Big]^2{{\Gamma({1\over
2}+\beta-ik)}\over{\Gamma({1\over 2}-\beta-ik)\Gamma({1\over
2}-\beta-\kappa)}}\Big[8iMr_+\sqrt{\mu^2-\omega^2}(m\Omega_{\text{H}}-\omega)\Big]^{2\beta}\
\end{equation}
for the bound-state resonances of the composed
Kerr-black-hole-massive-scalar-field system. Here
\begin{equation}\label{Eq12}
\Omega_{\text{H}}={{a}\over{r^2_++a^2}}
\end{equation}
is the angular velocity of the black-hole horizon, and
\begin{equation}\label{Eq13}
k\equiv 2\omega r_+\ \ \ \ , \ \ \ \ \kappa\equiv {{\omega
k-\mu^2r_+}\over{\sqrt{\mu^2-\omega^2}}}\ \ \ \ , \ \ \ \
\beta^2\equiv a^2\omega^2-2ma\omega+\mu^2r^2_++A_{lm}-k^2+{1\over
4}\ ,
\end{equation}
where $\{A_{lm}\}$ are the angular eigenvalues which couple the
radial Teukolsky equation (\ref{Eq9}) to the angular (spheroidal)
equation (\ref{Eq8}). These angular eigenvalues can be expanded in
the form \cite{Abram}
\begin{equation}\label{Eq14}
A_{lm}=l(l+1)+\sum_{k=1}^{\infty}c_ka^{2k}(\mu^2-\omega^2)^k\  ,
\end{equation}
where the expansion coefficients $\{c_k\}$ are given in
\cite{Abram}.

Before proceeding, it should be emphasized that the validity of the
resonance equation (\ref{Eq11}) is restricted to the regime
\begin{equation}\label{Eq15}
\tau\ll M(m\Omega_{\text{H}}-\omega)\ll x_{\text{o}}\ll
{{1}\over{M\sqrt{\mu^2-\omega^2}}}\  ,
\end{equation}
where $\tau\equiv (r_+-r_-)/r_+\ll1$ is the dimensionless
temperature of the rapidly-rotating (near-extremal) Kerr black hole,
and the dimensionless coordinate $x_{\text{o}}\equiv
(r_{\text{o}}-r_+)/r_+$ belongs to the {\it overlap} region in which
the two different solutions of the radial Teukolsky equation
(hypergeometric and confluent hypergeometric radial wave functions)
can be matched together, see \cite{HodSO,Ros} for details. The
inequalities in (\ref{Eq15}) imply that the resonance condition
(\ref{Eq11}) should be valid in the regime \cite{Noteinq}
\begin{equation}\label{Eq16}
M^2(m\Omega_{\text{H}}-\omega)\sqrt{\mu^2-\omega^2}\ll1\ .
\end{equation}

\section{The quasi-stationary bound-state resonances of the composed
black-hole-massive-scalar-field system}

As we shall now show, the resonance condition (\ref{Eq11}) can be
solved {\it analytically} in the physical regime (\ref{Eq16}). In
particular, in the present section we shall derive a (remarkably
simple) analytical formula for the discrete spectrum of oscillation
frequencies, $\{\omega_{\text{R}}(l,m,\alpha;n)\}_{n=0}^{n=\infty}$,
which characterize the quasi-stationary bound-state resonances of
the composed Kerr-black-hole-massive-scalar-field system.

Our analytical approach is based on the fact that the
right-hand-side of the resonance equation (\ref{Eq11}) is small in
the regime (\ref{Eq16}) with $\beta\in\mathbb{R}$ \cite{Notebr}. The
resonance condition can therefore be approximated by the simple
zeroth-order equation
\begin{equation}\label{Eq17}
{1\over{\Gamma({1\over 2}+\beta-\kappa)}}=0\  .
\end{equation}
As we shall now show, this zeroth-order resonance condition can be
solved analytically to yield the real oscillation frequencies which
characterize the bound-state scalar resonances. We first use the
well-known pole structure of the Gamma functions \cite{Abram} in
order to write the resonance equation (\ref{Eq17}) in the form
\cite{HodSO}
\begin{equation}\label{Eq18}
{1\over 2}+\beta-\kappa=-n  ,
\end{equation}
where the integer $n=0,1,2,...$ is the resonance parameter of the
field mode.

Defining the dimensionless variable \cite{Noteeps}
\begin{equation}\label{Eq19}
\epsilon\equiv M\sqrt{\mu^2-\omega^2}\  ,
\end{equation}
one finds from Eq. (\ref{Eq13})
\begin{equation}\label{Eq20}
\beta^2=\beta^2_0+O(\epsilon^2,\tau)\ \ \ \text{and}\ \ \
\kappa={{\alpha^2}\over{\epsilon}}-2\epsilon+O(\tau)\ ,
\end{equation}
where
\begin{equation}\label{Eq21}
\beta^2_0\equiv(l+1/2)^2-2m\alpha-2\alpha^2\ .
\end{equation}
Substituting (\ref{Eq20}) into the resonance condition
$\beta^2=[\kappa-(n+1/2)]^2$ [see Eq. (\ref{Eq18})], one obtains the
characteristic equation
\begin{equation}\label{Eq22}
\epsilon^2\cdot[(2l+1)^2-8m\alpha+8\alpha^2-(2n+1)^2]+\epsilon\cdot
4(2n+1)\alpha^2-4\alpha^4+O(\tau,\epsilon^3)=0\
\end{equation}
for the dimensionless parameter $\epsilon$. This resonance equation
can easily be solved to yield
\begin{equation}\label{Eq23}
\epsilon(l,m;n)={{2\alpha^2}\over{\sqrt{(2l+1)^2-8m\alpha+8\alpha^2}+1+2n}}\
.
\end{equation}

Finally, taking cognizance of the relation (\ref{Eq19}), one finds
the discrete spectrum of oscillation frequencies
\begin{equation}\label{Eq24}
{{\omega_{\text{R}}(n)}\over{\mu}}=\sqrt{1-\Big({{\alpha}\over{\ell+1+n}}\Big)^2}
\end{equation}
which characterize the quasi-stationary bound-state resonances of
the composed black-hole-massive-scalar-field system. Here
\begin{equation}\label{Eq25}
\ell\equiv {1\over 2}\big[\sqrt{(2l+1)^2-8m\alpha+8\alpha^2}-1\big]\
\end{equation}
is the generalized (finite-mass) spheroidal harmonic index. Note
that $\ell\to l$ in the small mass $\alpha\ll1$ limit, in which case
one recovers from (\ref{Eq24}) the well-known hydrogenic spectrum
(\ref{Eq4}) of \cite{Det}.

It is worth noting that, in general, the parameter $\ell(\alpha)$ is
not an integer. This implies that, for generic values of the
dimensionless mass parameter $\alpha$, the non-hydrogenic
oscillation spectrum (\ref{Eq24}) is {\it not} degenerate
\cite{Notecom}.

\section{Numerical confirmation}

It is of physical interest to test the accuracy of the analytically
derived formula (\ref{Eq24}) for the characteristic oscillation
frequencies $\omega^{\text{ana}}_{\text{R}}(n)/\mu$ of the
quasi-stationary massive scalar configurations. The quasi
bound-state resonances can be computed using standard numerical
techniques, see \cite{Dol1,Dol2} for details. In Table \ref{Table1}
we present a comparison between the analytically derived oscillation
frequencies (\ref{Eq24}) and the numerically computed resonances
\cite{Dol2}. The data presented is for the fundamental $l=m=1$ mode
with $a/M=0.99$ \cite{Notebg} and $\alpha=0.42$
\cite{Notefn,Notelm1}. One finds
a good agreement
between the {\it analytically} calculated oscillation frequencies
(\ref{Eq24}) and the {\it numerically} computed resonances of
\cite{Dol2}.

\begin{table}[htbp]
\centering
\begin{tabular}{|c|c|c|c|c|c|}
\hline
\ \ \text{Resonance parameter} $n$\ \ & \ 0\ \ & \ \ 1\ \ & \ \ 2\ \ & \ \ 3\ \ & \ \ 4\ \ \\
\hline \ \
$\omega^{\text{ana}}_{\text{R}}(n)/\omega^{\text{num}}_{\text{R}}(n)$\
\ & \ \ 0.9999\ \ \ & \ \ 1.0006\ \ \ & \ \ 1.0004\ \ \ &
\ \ 1.0002\ \ \ & \ \ 0.9994\ \ \ \\
\hline
\end{tabular}
\caption{Quasi-stationary resonances of massive scalar fields in the
rotating Kerr black-hole spacetime. The data shown is for the
fundamental $l=m=1$ mode with $a/M=0.99$, $\alpha\equiv M\mu=0.42$,
and $n=0,1,2,3,4$. We display the dimensionless ratio between the
{\it analytically} derived oscillation frequencies
$\omega^{\text{ana}}_{\text{R}}(n)$ [see Eq. (\ref{Eq24})] and the
{\it numerically} computed resonances
$\omega^{\text{num}}_{\text{R}}(n)$ of \cite{Dol2}. One finds
a good agreement between the analytical formula (\ref{Eq24}) and the
numerical data of \cite{Dol2}
.} \label{Table1}
\end{table}

In order to compare the accuracy of the newly derived analytical
formula (\ref{Eq24}) with the accuracy of the familiar hydrogenic
(small-mass) spectrum (\ref{Eq4}), we display in Table \ref{Table2}
the physical quantity $\epsilon(n)$ [see Eq. (\ref{Eq19})] which
provides a quantitative measure for the deviation of the resonant
oscillation frequency $\omega_{\text{R}}(n)$ from the field mass
parameter $\mu$. In particular, we present the dimensionless ratios
$\epsilon^{\text{ana}}/\epsilon^{\text{num}}$ and
$\epsilon^{\text{ana-hydro}}/\epsilon^{\text{num}}$, where
$\epsilon^{\text{ana}}(n)$ is given by the analytical formula
(\ref{Eq24}), $\epsilon^{\text{ana-hydro}}(n)$ is defined from the
hydrogenic spectrum (\ref{Eq4}), and $\epsilon^{\text{num}}(n)$ is
obtained from the numerically computed resonances of \cite{Dol2}.
One finds that, in general, the newly derived formula (\ref{Eq24})
performs better than the hydrogenic formula (\ref{Eq4})
\cite{Noteexc}.

\begin{table}[htbp]
\centering
\begin{tabular}{|c|c|c|c|c|c|}
\hline
\ \ \text{Resonance parameter} $n$\ \ & \ 0\ \ & \ \ 1\ \ & \ \ 2\ \ & \ \ 3\ \ & \ \ 4\ \ \\
\hline \ \ $\epsilon^{\text{ana}}(n)/\epsilon^{\text{num}}(n)$\ \ &
\ \ 1.001\ \ \ & \ \ 0.974\ \ \ & \ \ 0.971\ \ \ &
\ \ 0.974\ \ \ & \ \ 1.140\ \ \ \\
\hline \ \
$\epsilon^{\text{ana-hydro}}(n)/\epsilon^{\text{num}}(n)$\ \ & \ \
0.910\ \ \ & \ \ 0.916\ \ \ & \ \ 0.928\ \ \ &
\ \ 0.939\ \ \ & \ \ 1.107\ \ \ \\
\hline
\end{tabular}
\caption{Quasi-stationary resonances of massive scalar fields in the
rotating Kerr black-hole spacetime. The data shown is for the
fundamental $l=m=1$ mode with $a/M=0.99$, $\alpha\equiv M\mu=0.42$,
and $n=0,1,2,3,4$. We display the dimensionless ratios
$\epsilon^{\text{ana}}/\epsilon^{\text{num}}$ and
$\epsilon^{\text{ana-hydro}}/\epsilon^{\text{num}}$, where
$\epsilon^{\text{ana}}(n)$ is given by the analytical formula
(\ref{Eq24}), $\epsilon^{\text{ana-hydro}}(n)$ is defined from the
hydrogenic spectrum (\ref{Eq4}), and $\epsilon^{\text{num}}(n)$ is
obtained from the numerically computed resonances of \cite{Dol2}.
One finds that, in general, the newly derived analytical formula
(\ref{Eq24}) performs better than the hydrogenic formula (\ref{Eq4})
\cite{Noteexc}.} \label{Table2}
\end{table}

\section{Summary}

In summary, we have studied the resonance spectrum of
quasi-stationary massive scalar configurations linearly coupled to a
near-extremal (rapidly-rotating) Kerr black-hole spacetime. In
particular, we have derived a compact analytical expression [see Eq.
(\ref{Eq24})] for the characteristic oscillation frequencies
$\omega^{\text{ana}}_{\text{R}}(n)/\mu$ of the bound-state massive
scalar fields. It was shown that the {\it analytically} derived
formula (\ref{Eq24}) agrees with direct {\it numerical} computations
\cite{Dol2} of the black-hole-scalar-field resonances.

It is well known that the characteristic hydrogenic spectrum
(\ref{Eq4}) in the {\it small}-mass $\alpha\ll1$ limit is highly
degenerate -- it depends only on the principal resonance parameter
$\bar n_0\equiv l+1+n$ \cite{Noteint} [Thus, according to
(\ref{Eq4}), two different modes which are characterized by the
integer parameters $(l,n)$ and $(l',n')$ with $l+n=l'+n'$ share the
{\it same} resonant frequency $\omega_{\text{R}}(\bar n_0)$ in the
$\alpha\to 0$ limit]. On the other hand, the newly derived resonance
spectrum (\ref{Eq24}), which is valid in the $\alpha=O(1)$ regime,
is no longer degenerate. That is, for generic values of the
dimensionless mass parameter $\alpha$, two quasi-stationary modes
with different sets of the integer parameters $(l,n)$ are
characterized, according to (\ref{Eq24}), by {\it different}
oscillation frequencies \cite{Notenil}.

Finally, it is worth emphasizing again that the physical
significance of the characteristic oscillation frequencies
(\ref{Eq24}) lies in the fact that these quasi-stationary
(long-lived) resonances dominate the dynamics [and, in particular,
dominate the characteristic Fourier power spectra $P(\omega)$
\cite{Dol2}] of the massive scalar fields in the black-hole
spacetime.

\bigskip
\noindent
{\bf ACKNOWLEDGMENTS}
\bigskip

This research is supported by the Carmel Science Foundation. I would
like to thank Yael Oren, Arbel M. Ongo, Ayelet B. Lata, and Alona B.
Tea for helpful discussions.


\begin{thebibliography}{99}
\bibitem{Noteunits} We use natural units in which $G=c=\hbar=1$.
\bibitem{Notedim} Note that the field mass parameter $\mu$ stands for $\mu/\hbar$. Hence, it
has the dimensions of $($length$)^{-1}$.
\bibitem{Notenn} Here the integer $n=0,1,2,...$ is the resonance
parameter of the radial field mode, see Eq. (\ref{Eq18}) below.
\bibitem{Notefor} More explicitly, by recording the temporal dependence of the scalar-field
amplitude $A(t)$ at some fixed point $r=r_0$, and taking the Fourier
transform of this time-dependent field amplitude, one finds
\cite{Dol2} that the resulting Fourier power spectrum $P(\omega)$ is
characterized by sharp picks at the appropriate resonant frequencies
$\omega_{\text{R}}(\mu)$ [see Eq. (\ref{Eq24}) below] of the
composed black-hole-scalar-field system (see, in particular, Fig. 12
of \cite{Dol2}). This implies that the quasi-stationary bound-state
scalar resonances are excited by generic initial data of the massive
scalar fields.
\bibitem{Notemn} Here $\omega, l$, and $m$ are respectively the (conserved) frequency of the field mode,
its spheroidal harmonic index, and its azimuthal harmonic index.
\bibitem{Notepm} Here $r_{\pm}=M\pm (M^2-a^2)^{1/2}$ are the
black-hole (event and inner) horizons.
\bibitem{Notetor} Here the ``tortoise" radial coordinate $y$ is defined
by the relation $dy=[(r^2+a^2)/\Delta]dr$.
\bibitem{Noteinq} In particular, since each inequality in (\ref{Eq15}) roughly corresponds to an
order-of-magnitude difference between two quantities, one concludes
that the resonance condition (\ref{Eq11}) should be valid in the
regime $M^2(m\Omega_{\text{H}}-\omega)\sqrt{\mu^2-\omega^2}\lesssim
10^{-2}$.
\bibitem{Notebr} One finds that $\beta\in\mathbb{R}$ in the regime
$\alpha\leq [\sqrt{2(l+1/2)^2+m^2}-m]/2$, see Eq. (\ref{Eq21})
below.
\bibitem{Notecom} Compare this property of the non-hydrogenic spectrum (\ref{Eq24})
with the familiar hydrogenic spectrum (\ref{Eq4}) which is known to
be {\it degenerate}.
\bibitem{Notebg} This is the largest black-hole spin for which we have exact (numerical) data \cite{Dol2} for the
characteristic oscillation frequencies $\omega_{\text{R}}(n)/\mu$ of
the quasi-stationary bound-state resonances. Note that the numerical
results of \cite{Dol1} indicate that, for $\alpha\lesssim0.6$ and
$a\to1^{-}$, the values of the fundamental resonant frequencies
$\omega_{\text{R}}/\mu$ depend very weakly on the black-hole
rotation parameter $a$ (see, in particular, Fig. 5 of \cite{Dol1}).
\bibitem{Notefn} As discussed above, our resonance equation (\ref{Eq11}) is restricted to the regime
$M^2(m\Omega_{\text{H}}-\omega)\sqrt{\mu^2-\omega^2}\lesssim10^{-2}$
\cite{Noteinq}. Substituting our solution (\ref{Eq24}) into the
left-hand-side of (\ref{Eq16}), one finds that, for the fundamental
$l=m=1$ mode, the requirement
$M^2(m\Omega_{\text{H}}-\omega)\sqrt{\mu^2-\omega^2}\lesssim10^{-2}$
is respected in the regime $\alpha\lesssim 0.58$. Note that this
mass regime is consistent with our previous requirement
$\beta\in\mathbb{R}$ \cite{Notebr}, which implies the weaker
restriction $\alpha\lesssim 0.67$ for the fundamental $l=m=1$ mode.
\bibitem{Noteexc} The only exception is the excited $n=4$ mode, for
which the agreement between the numerically computed
deviation-parameter $\epsilon^{\text{num}}(n=4)$ and the analytical
formulas (\ref{Eq4}) and (\ref{Eq24}) is quite poor.
\bibitem{Noteint} Note that this parameter is an integer.
\bibitem{Notenil} Note that, for generic values of the mass parameter $\alpha$,
the generalized (finite-mass) spheroidal harmonic index
$\ell(\alpha)$ that appears in the resonance spectrum (\ref{Eq24})
is not an integer.
\end{thebibliography}
\end{document}